\documentclass{article}

\usepackage{arxiv}

\usepackage[utf8]{inputenc} 
\usepackage[T1]{fontenc}    
\usepackage{hyperref}       
\usepackage{url}            
\usepackage{booktabs}       
\usepackage{amsfonts}       
\usepackage{nicefrac}       
\usepackage{microtype}      
\usepackage{lipsum}
\usepackage{bm}
\usepackage{amsmath}
\usepackage{graphicx}

\title{Dynamics of collision of a spherical ice crystal particle with  a perfectly rigid substrate}

\author{
Ilia V. Roisman \\
Institute for Fluid Mechanics and Aerodynamics, Technische Universität Darmstadt, Darmstadt, Germany\\
  \texttt{roisman@sla.tu-darmstadt.de} \\
}

\begin{document}
\maketitle
\begin{abstract}
Impact of single particle onto a rigid substrate leads to its deformation and fragmentation. The flow associated with the particle spreading on a solid substrate after impact is extremely complicated. In this theoretical study a simplified model for the plastic  flow with the rate-dependent yield strength is developed. The flow in the particle is approximated by an incompressible inviscid flow past a thin rigid disk. The expression for the pressure field distribution is obtained in the vicinity of the impact axis. The total momentum balance of the particle is used to derive the equations of the particle deformation by impact. The theoretical predictions of the typical geometrical parameters of the particle, the peak force and the evolution of the force in time are compared with the existing experimental data. The agreement is rather good.
\end{abstract}


\section{Introduction}

Ice particle impact onto a  solid substrate is influenced by the impact parameters and material properties of the target. At relatively  low  velocities the impact leads to the elastic rebound. Starting from some limiting velocity the collision leads to a significant particle plastic deformation and even break-up. If the target or the particle is wet the impact could lead to the particle deposition and further agglomeration.

Important industrial  processes are associated with single ice particle impact. One of them is the ice crystal ice accretion in the low pressure compression system of an aircraft \cite{currie2014experimental,mason2011understanding,currie2012fundamental,bucknell2018experimental,saunders1998laboratory,veres2013modeling,ayan2015prediction,veres2012model} where the walls are hot enough to partially melt the particles during collision. Ice crystal accretion on internal components of aircraft engines and further ice layer shedding \cite{mason2011understanding,mazzawy2007modeling} may lead to flameout, mechanical damage, rollback, etc. Sensors on the aircraft can also be affected by
ice crystal accretion.

The physics of ice crystal accretion is not yet completely understood. It is known that the inception of accretion can be associated with the cooling of an initially hot substrate. As soon as the wall exposed to the flow of cold ice crystals reaches the melting temperature the ice layer starts to grow by accumulating of the fragments of the impacted particles \cite{lowe2016inception,hauk2016investigation}. Another important reason leading to the particle deposition is the capillary forces associated with the liquid bridges formed between wet contacting particles. In the case of impacting particle the liquid bridge may decelerate the rebounding particle significantly \cite{antonyuk2009influence,cruger2016coefficient,gollwitzer2012coefficient,mullins2003particle,fu2005experimental}. The particle will agglomerate on the substrate if the typical particle deposition time, determined by the viscous, inertial and capillary forces in the liquid bridge, is shorter than the breakup time of the liquid bridge \cite{brulin2020pinch,weickgenannt2015pinch,qian2011motion,curran2005liquid}.

Ice crystal impact studies are also motivated by the atmospheric and  meteorological sciences, since ice collision phenomena influence the physical processes in clouds, thunderstorm electrification processes,\cite{saunders1993review,brooks1997effect,saunders1998laboratory} can lead to the generation of fine crystal particles in clouds \cite{vardiman1978generation,murphy2004particle}.

In general particle collision with solid surfaces and between each other is a phenomenon relevant to many other technological processes, mainly governed by the dynamics of granular media. Among the examples of such applications are mineral processing, agricultural products,
detergents, pharmaceuticals, food products and chemicals. The breakage of the particles leads to the evolution of the size distribution of the particles \cite{reynolds2005breakage}, while coalescence (promoted by a presence of a liquid phase), or sintering in some cases, can cause a catastrophic defluidisation \cite{geldart1972effect,antony2004granular}.
Collision phenomena,  including coagulation, rebound, and fragmentation, are also important physical processes controlling the dynamics of Saturn's rings \cite{higa1998size,supulver1995coefficient} or formation processes of satellites and planets in the outer Solar System \cite{kato1995ice}.


Many existing theoretical approaches to the modeling of the impact, deformation and penetration of solid bodies are based on the hydrodynamic theories \cite{rosenberg1990hydrodynamic,tate1986long,frankel1990hydrodynamic,lou2014long,seguin2011dense,anderson2017analytical,yarin2017collision}. Such models are often based on a kinematically admissible flow, which satisfies the continuity equations, the boundary conditions and the momentum balance conditions in a simplified form, for example in integral form of in certain regions of the deforming media. Such models are able to successfully predict integral parameters of the problem, like penetration depth, projectile deformation, its residual velocity after penetration of a finite target.

Approximate model of deformation of a long plastic rod has been developed in \cite{taylor1948use}. The model is based on the assumption that the rod is deformed by the propagation of a plastic wave, generated by impact onto a perfectly rigid target. Similar ideas have been applied in \cite{Hauk.2015,roisman2015impact} for the description of deformation and breakup of an ice particle. However, these simplified approaches are not sufficient to model the processes of particle fragmentation or heat transfer. For such advanced modeling some details of the velocity field in the deforming particle are required \cite{yarin2000model}.

The main objective of this study is the theoretical model for deformation of a plastic particle during its collision with a perfectly rigid target. The flow in the particle is approximated by an instantaneous potential flow past a disc. The  disc in the model is associated with the circular spot of the deforming particle and the rigid substrate. In the plastic flow in the particle the yields strength $Y$ is a function of the effective strain rate. Such material behaviour is typical to quasi-brittle ice crystal particles \cite{tippmann2013experimentally}. The momentum balance equation is solve in the proximity to the particle impact axis. The  stress field in the deforming particle allows to estimate the total contact force and to predict evolution of the particle deformation. The predicted evolution of the force applied to the target by an impacting ice particle agrees well with the experimental data existing in the literature.

\section{Flow field in the deforming particle}

The flow field in a  particle deforming by a collision with a rigid target is determined by the material properties of the ice, like yield strength $Y$, elastic shear modulus $G$, density. The flow can be also influenced by the developments of cracks leading to the particle fragmentation. Nevertheless, in many cases a simplified kinematically admissible flow  can be a rather good approximation for the velocity field in the particle, especially if the inertial effects in the flow are dominant. Such approach is useful for the description of the velocity fields in the penetration mechanics \cite{yarin2017collision}. In this study a flow in the deforming particle is approximated by an incompressible potential flow, associated with the axisymmetric flow around a thin disc. The assumption of the particle incompressibility is valid for the impact velocities much smaller that the speed of sound in ice. The speed of sound for pressure waves at $-10 ^\circ$C is 3865 m/s, which is much higher than the typical impact velocities of 100 m/s relevant to the aircraft applications.

Consider a cylindrical coordinate system $\{r,\theta,z\}$ with the corresponding base unit vectors $\{\bm e_r, \bm e_\theta, \bm e_z\}$.  The flow field in the deforming particle is approximated by the flow around a flat disc of the radius $a$, where $a$ is the instantaneous  radius of the impression. The velocity potential for this flow relative to the wall is obtained from the well-known solution \cite{batchelor1967introduction} in the form
\begin{equation}\label{eq_phi}
    \phi = \frac{2 a U}{\pi} \cos\eta \left[\sinh\xi \cot^{-1}(\sinh\xi) -1\right] - U z,
\end{equation}
where $\xi, \eta$ are dimensionless elliptic coordinates defined through
\begin{equation}\label{xieta}
    \xi + i \eta= \sinh^{-1}\left(\frac{z+i r}{a}\right).
\end{equation}


This velocity satisfies the impenetrability conditions at the solid substrate $z=0, r < a$. The velocity field at the infinity approaches to the uniform flow $u_z=-U$.

Equation (\ref{xieta}) allows to explicitly derive the expressions for the coordinates in the cylindrical system
\begin{subequations}
\begin{eqnarray}
    \xi &=& \frac{1}{2} \ln \left[\frac{(z+S \cos \beta)^2+(r+S \sin \beta)^2}{a^2} \right]\\
    \eta &=& \tan^{-1}\left[\frac{r+S \sin \beta}{z+S \cos \beta}\right]\\
     S&=& \left[4 r^2 z^2 +(a^2-r^2+z^2)^2\right]^{1/4},\\
    \beta &=& \left\{\begin{matrix}
\frac{1}{2} \tan^{-1}\left(\frac{2 r z}{z^2+a^2-r^2}\right) & \mathrm{if}\, r^2< z^2+a^2 \\
\frac{1}{2} \tan^{-1}\left(\frac{2 r z}{z^2+a^2-r^2}\right)+\frac{\pi}{2} & \mathrm{if}\, r^2 >z^2+a^2\\
\frac{\pi}{2} & \mathrm{if}\, r^2 =z^2+a^2
\end{matrix}\right.\label{eq_beta}
\end{eqnarray}
\end{subequations}

The components of the velocity field in the cylindrical coordinate system can be now obtained from the potential function
\begin{equation}\label{eq_u}
    \bm u = \bm\nabla \phi= (\phi_{,\xi} \xi{,_r} + \phi_{,\eta} \eta_{,r})\bm e_r + (\phi_{,\xi} \xi{,_z} + \phi_{,\eta} \eta_{,z})\bm e_z.
\end{equation}

\subsection{Velocity field in the vicinity of the impact axis}
The velocity components can be obtained from (\ref{eq_u}) with the help of (\ref{eq_phi})-(\ref{eq_beta}). Near the axis the expressions for the velocity components can be reduced to
\begin{subequations}
\begin{eqnarray}
    u_r & =& U\left[\frac{2 a^3 r }{\pi  \left(a^2+z^2\right)^2} +   \mathcal{O}\left(\frac{r^3}{a^3}\right)\right]\\
    u_z &= &U\left[\frac{2 \cot ^{-1}\left(\frac{z}{a}\right) }{\pi }-\frac{2 a z }{\pi
   \left(a^2+z^2\right)}-1 -\frac{4 a^3 r^2  z}{\pi  \left(a^2+z^2\right)^3} +  \mathcal{O}\left(\frac{r^3}{a^3}\right)\right] \label{eq:uz}
\end{eqnarray}
\end{subequations}

The corresponding  rate-of-strain tensor $\bm E$ is the symmetric part of the velocity gradient at the impact axis
\begin{eqnarray}
    \bm E &=& \frac{\dot\gamma}{2} \left(\bm e_r \otimes \bm e_r-\frac{4r z}{a^2+z^2}(\bm e_r \otimes \bm e_z+\bm e_z \otimes \bm e_r) + \bm e_\theta \otimes \bm e_\theta - 2 \bm e_z \otimes \bm e_z\right),\label{eq:strainr}\\
    \dot\gamma&=&\frac{4 a^3 U}{\pi  \left(a^2+z^2\right)^2},\label{eq:strainrateeq}
\end{eqnarray}
where the symbol $\otimes$ denotes the usual tensor product, and $\dot\gamma \equiv \sqrt{2/3}\sqrt{\bm E:\bm E}$ is the equivalent rate of strain.

\subsection{Dimensionless particle dislodging during impact}
Let us introduce a dimensionless particle dislodging $\zeta(t)$ scaled by the particle initial radius $R$
\begin{equation}
    \zeta = \frac{1}{R} \int_0^t U(t) \mathrm{d}t.
\end{equation}


The evolution of the particle height can be estimated by the numerical integration of the equation $\mathrm{d}h/\mathrm{d}t = u_z(z=h)$ leading to
\begin{subequations}\label{eq:difh}
\begin{eqnarray}
  \frac{1}{R}  \frac{\mathrm{d} h}{\mathrm{d} \zeta}& =& \frac{2 \cot ^{-1}\left(\frac{h}{a}\right)}{\pi }-\frac{2 a h }{\pi\left(a^2+h^2\right)}-1,\\
    h &=& 2 R, \quad \mathrm{at}\quad \zeta = 0.
\end{eqnarray}
\end{subequations}

\begin{figure}
\centering
\includegraphics[width=0.6\textwidth]{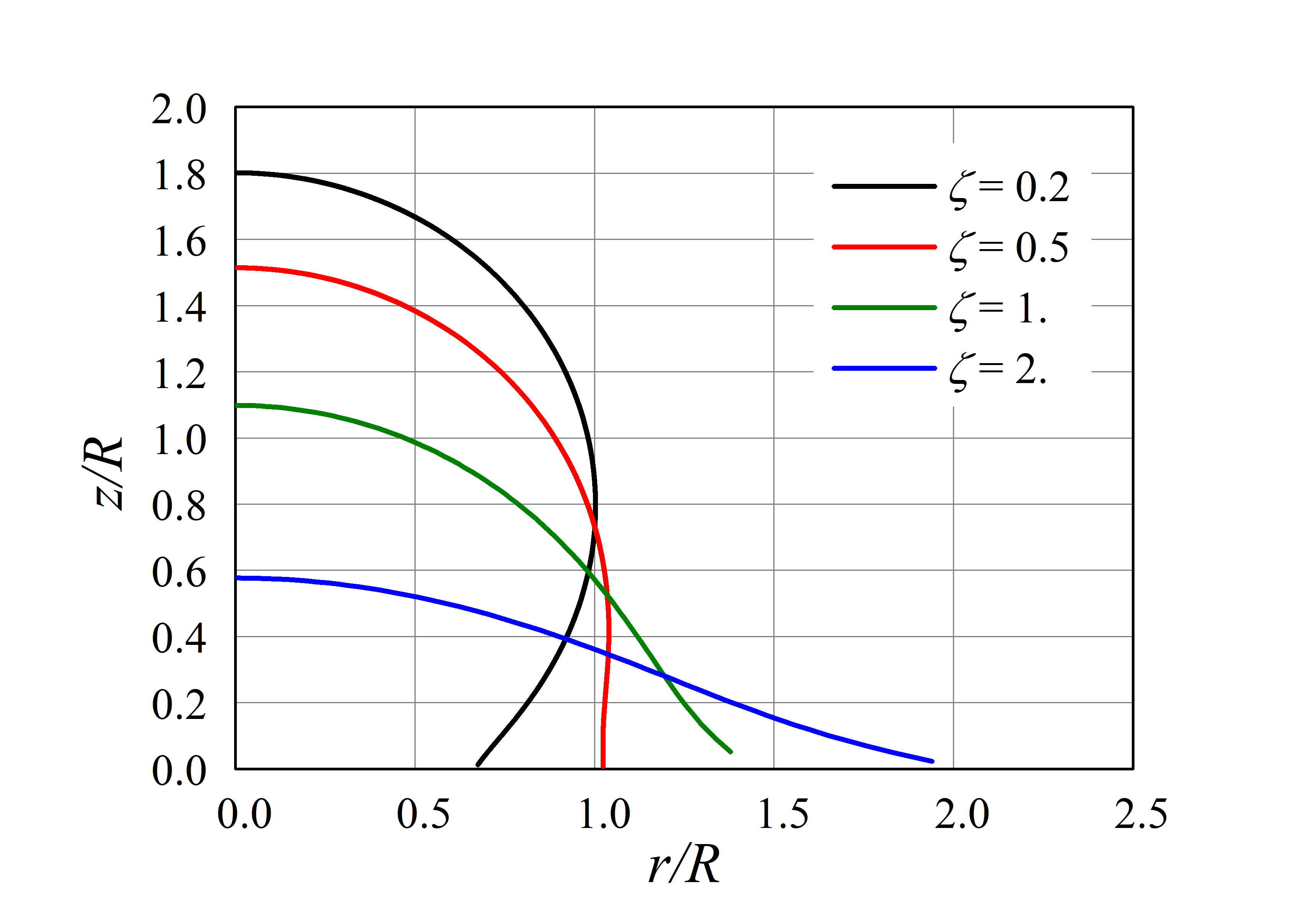}
\caption{Theoretically predicted dimensionless shapes of the impacting particle at different values of the dimensionless displacement $\zeta$.}
\label{fig:shapes}
\end{figure}

It is known that in the case of impact of an inviscid drop the flow near the wall does not influence the outer flow. The value of $U$ is thus constant and the dislodging is reduced to the dimensionless time $\zeta = U_0/R$, where $U_0$ is the impact velocity. Theoretically predicted shapes  of the impacting particle, computed using the velocity field (\ref{eq_u}), is shown in Fig.~\ref{fig:shapes}. These predictions do not account for the ejection of the lamella along the substrate. Nevertheless, the shapes are very similar to the observed forms of the deforming liquid drop during its high-velocity impact onto a dry substrate.

\begin{figure}
\centering
\includegraphics[width=0.6\textwidth]{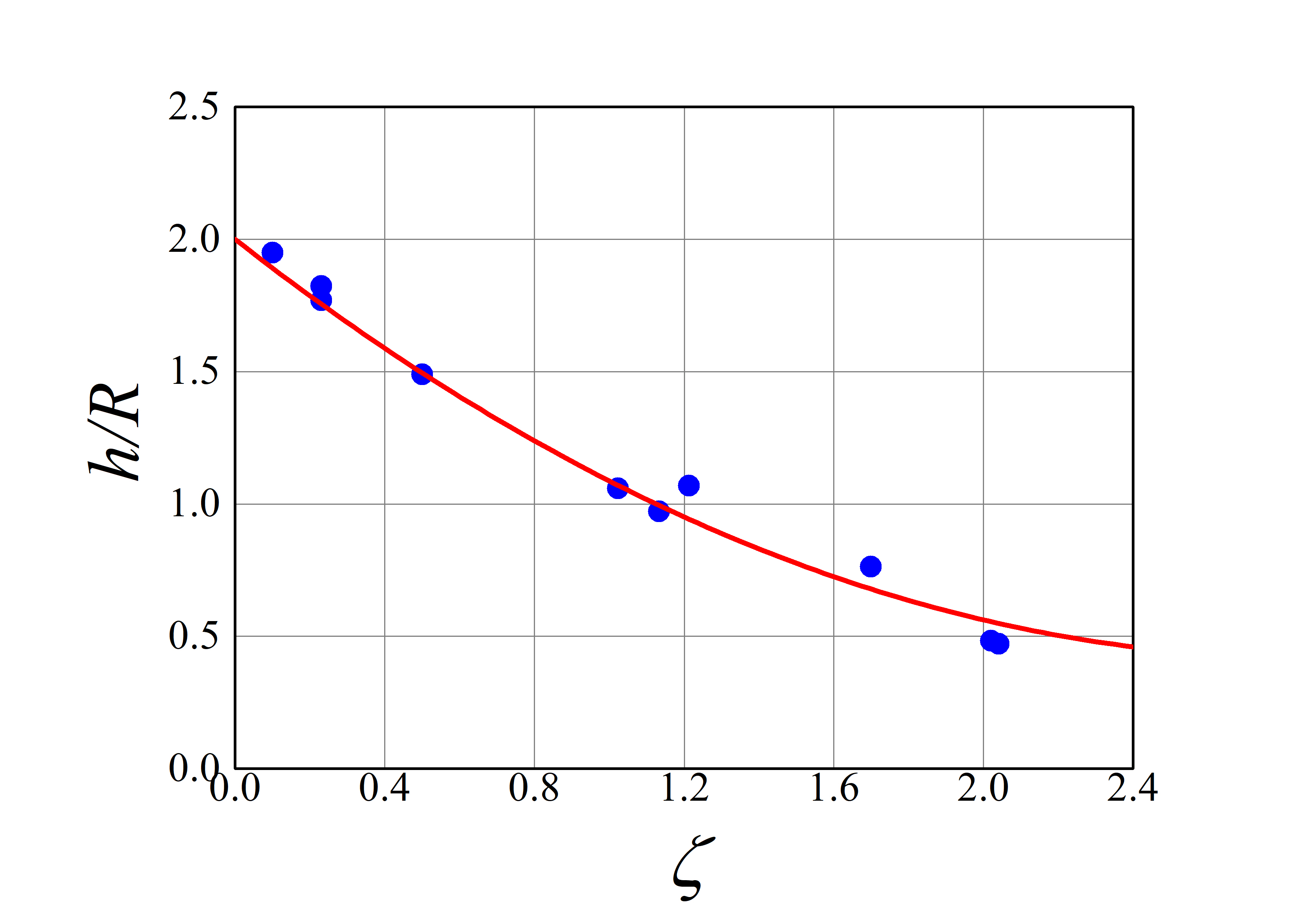}
\caption{Theoretically predicted dimensionless particle height $h/R$ as a function of $\zeta$, in comparison with the computations for high-speed impact of a liquid drop \cite{roisman2009inertia}.}
\label{fig:Height}
\end{figure}

In Fig.~\ref{fig:Height} the predicted particle dimensionless height $h/R$ is shown as a function of $\zeta$ in comparison with the computations of a liquid drop impacts with very high Reynolds and Weber numbers \cite{roisman2009inertia}. These computations are validated by comparison with the numerous experimental data. The agreement between the present theory and the CFD computations is rather good, although no adjustable parameters have been introduced in the theory.

\section{Stress fields in the deforming plastic particle}
\subsection{Plastic stresses near the particle axis}
The stress tensor in the particle is determined by the velocity field and by the yield strength $Y$
\begin{equation}\label{eq:sigPL}
    \bm \sigma = - p \bm I +\bm \sigma',\quad \bm\sigma'= \left(\frac{2}{3}\right)^{1/2} \frac{Y}{\sqrt{\bm E:\bm E}} \bm E
\end{equation}
where $p$ is the pressure and $\bm I$ is the unit tensor.

In this study the effect of the strain rate on the hardening of the particle material is taken into account. The material hardening is an important property of the ice crystals \cite{tippmann2013experimentally}. This means the uniaxial yield strength is modeled in the form:
\begin{equation}\label{formY}
    Y = Y_0 y(\dot\gamma),
\end{equation}
where $y$ is a dimensionless function of the the equivalent rate of strain $\dot\gamma$, determined in (\ref{eq:strainrateeq}) and $Y_0$ is the static yield strength at $\dot \gamma\rightarrow 0$.

Near the particle axis, $r\rightarrow 0$, the deviatoric part of the stress is obtained using (\ref{eq:strainr}) and (\ref{eq:sigPL}) and further linearization it for small $r$
\begin{equation}\label{sigmaStr}
    \bm\sigma' = \frac{Y}{3}  \left(\bm e_r \otimes \bm e_r-\frac{4r z}{a^2+z^2}(\bm e_r \otimes \bm e_z+\bm e_z \otimes \bm e_r) + \bm e_\theta \otimes \bm e_\theta - 2 \bm e_z \otimes \bm e_z\right).
\end{equation}

The momentum balance equation in the particle flow can be written in the form
\begin{equation}\label{Mom}
    \bm\nabla\left(\rho \frac{\partial \phi}{\partial t}+\frac{\rho}{2} \bm\nabla\phi\cdot\bm\nabla\phi + p \right) =  \bm\nabla\cdot\bm\sigma'.
\end{equation}
where the divergence of the deviatoric stress at $r\rightarrow 0$ is obtained from (\ref{sigmaStr})
\begin{equation}\label{divMom}
    \bm\nabla\cdot\bm\sigma' =\left[\frac{32  Y_0  a^3 U z}{3\pi  \left(a^2+z^2\right)^3} \frac{\mathrm{d} y(\dot\gamma)}{\mathrm{d}\dot\gamma} - \frac{8 z Y_0 }{3(a^2+z^2)}y(\dot\gamma)\right] \bm e_z.
\end{equation}

Integration of (\ref{Mom}) with the help of (\ref{divMom}) yields
\begin{equation}
    p = - \frac{2 Y}{3}+\frac{2 Y_0}{3} \int_{\dot\gamma(0)}^{\dot\gamma}\frac{y(\dot\gamma)}{\dot\gamma}\mathrm{d}\dot\gamma -\frac{\rho}{2} \bm\nabla\phi\cdot\bm\nabla\phi -\rho \frac{\partial \phi}{\partial t} + f(t),
\end{equation}
where $f(t)$ is a function of time which has to be determined from the boundary conditions. The expression for the pressure at the axis of the particle for the velocity potential defined in (\ref{eq_phi}) is therefore
\begin{eqnarray}
    p &=& f(t) - \frac{2 Y(z)}{3}+\frac{2 Y_0}{3} \int_{\dot\gamma(0)}^{\dot\gamma}\frac{y(\dot\gamma)}{\dot\gamma}\mathrm{d}\dot\gamma\nonumber \\
    &-& \frac{\rho z}{\pi }\left[\pi  -2  \cot ^{-1}\left(\frac{z}{a}\right)\right] \frac{\mathrm{d} U}{\mathrm{d} t} -\frac{2\rho z^2 U}{\pi  \left(a^2+z^2\right)} \frac{\mathrm{d} a}{\mathrm{d} t}\nonumber\\
    &-& \frac{\rho}{2} U^2 \left\{\frac{2 \cot ^{-1}\left(\frac{z}{a}\right)}{\pi }-\frac{2 a z }{\pi\left(a^2+z^2\right)}-1\right\}^2 \label{eq:p0t}
\end{eqnarray}

The function $f(t)$ is determined from the condition at the rear tip of the particle $z=h(t)$ where the normal stress $\sigma_{zz}= -p - 2 Y(h)/3$ vanishes:
\begin{eqnarray}
    f(t) &=& -\frac{2 Y_0}{3} \int_{\dot\gamma(0)}^{\dot\gamma(h)}\frac{y(\dot\gamma)}{\dot\gamma}\mathrm{d}\dot\gamma\\
    &+& \frac{\rho h}{\pi }\left[\pi  -2  \cot ^{-1}\left(\frac{h}{a}\right)\right] \frac{\mathrm{d} U}{\mathrm{d} t} +\frac{2\rho h^2 U}{\pi  \left(a^2+h^2\right)} \frac{\mathrm{d} a}{\mathrm{d} t} \nonumber\\
    &+& \frac{\rho}{2} U^2 \left\{\frac{2 \cot ^{-1}\left(\frac{h}{a}\right)}{\pi }-\frac{2 a h }{\pi\left(a^2+h^2\right)}-1\right\}^2 \label{eq:ft}
\end{eqnarray}

Finally, the normal stress at the wall surface, $z=0, \,r=0$, is obtained from (\ref{eq:p0t}) and (\ref{eq:ft}) in the form
\begin{subequations}
\begin{eqnarray}\label{eq:szzG}
 -\sigma_{zz} &=&  A(\zeta) \rho U\frac{\mathrm{d} U}{\mathrm{d} \zeta} + B(\zeta) \rho U^2 + C Y_0,  \label{eq:szz}\\
 A   &=& \frac{h}{R}\left[1  - \frac{2}{\pi}  \cot ^{-1}\left(\frac{h}{a}\right)\right] , \\
 B&=& \frac{2 h^2}{\pi R \left(a^2+h^2\right)} \frac{\mathrm{d} a}{\mathrm{d} \zeta}+ \frac{1}{2} \left\{\frac{2 \cot ^{-1}\left(\frac{h}{a}\right)}{\pi }-\frac{2 a h }{\pi\left(a^2+h^2\right)}-1\right\}^2, \label{eq:B}\\
  C &=& \frac{2}{3} \int_{\dot\gamma(h)}^{\dot\gamma(0)}\frac{y(\dot\gamma)}{\dot\gamma}\mathrm{d}\dot\gamma.\label{funC}
\end{eqnarray}
\end{subequations}
where $A(\zeta)$ and $B(\zeta)$ are dimensionless functions of the dimensionless particle dislodging $\zeta$, and $C$ is a dimensionless function based on the distribution of the of the strain rates.

\subsection{Momentum balance of an entire particle}

Let us assume the total axial momentum of the deforming particle in integral form
\begin{equation}\label{eq:defKY}
    P(\zeta) = \pi \rho R^3 U K(\zeta),
\end{equation}
where $K(\zeta)$ is a dimensionless function of $\zeta$. The momentum balance equation
\begin{equation}
    \frac{\mathrm{d}P}{\mathrm{d}t} \approx \pi a^2 \sigma_{zz},
\end{equation}
can be rewritten with the help of (\ref{eq:szzG}) and (\ref{eq:defKY}) in the form
\begin{eqnarray}
 \left(K + \frac{a^2 A}{R^2} \right)  U\frac{\mathrm{d}U}{\mathrm{d}\zeta}  &+&\left(\frac{\mathrm{d}K}{\mathrm{d}\zeta} + \frac{a^2 B}{R^2}\right)U^2 +  \frac{ a^2 C Y_0}{ R^2\rho} =0. \label{eqMomC}
\end{eqnarray}


The momentum balance equation can be solved numerically if the dependence of the dimensionless particle axial momentum $K(\zeta)$ is known. This function is determined from the consideration of a high-speed particle impact governed exclusively by the inertial terms.


The effect of the elastic region in the deforming particle is neglected in this study since the strains in the particle grow rather fast as soon the particle is deforming. The analysis of the elastic region is described in  \ref{appa}, where the minor effect of the elastic stresses is demonstrated.

\section{High impact velocity approximation in the limit $Y\ll \rho U^2$}
In the case of very high impact velocity the effect of the yield stress can be much smaller than the inertia. Therefore the terms associated with the yield strength in (\ref{eqMomC}) can be neglected.

Moreover, since the inviscid is flow is disturbed only in a finite region near the wall of the size comparable with $a$, the value of $U$ far from the wall is assumed to be constant. The following expression for the axial momentum balance is obtained from (\ref{eqMomC}) for  $U(\zeta)=U_0$
\begin{eqnarray}\label{eqforK}
    \frac{\mathrm{d}K}{\mathrm{d}\zeta} + \frac{a^2 B}{R^2}&=&0.\\
    \zeta &=& \frac{t U_0}{R}.
\end{eqnarray}

\begin{figure}
    \centering
    \includegraphics[width=0.6 \textwidth]{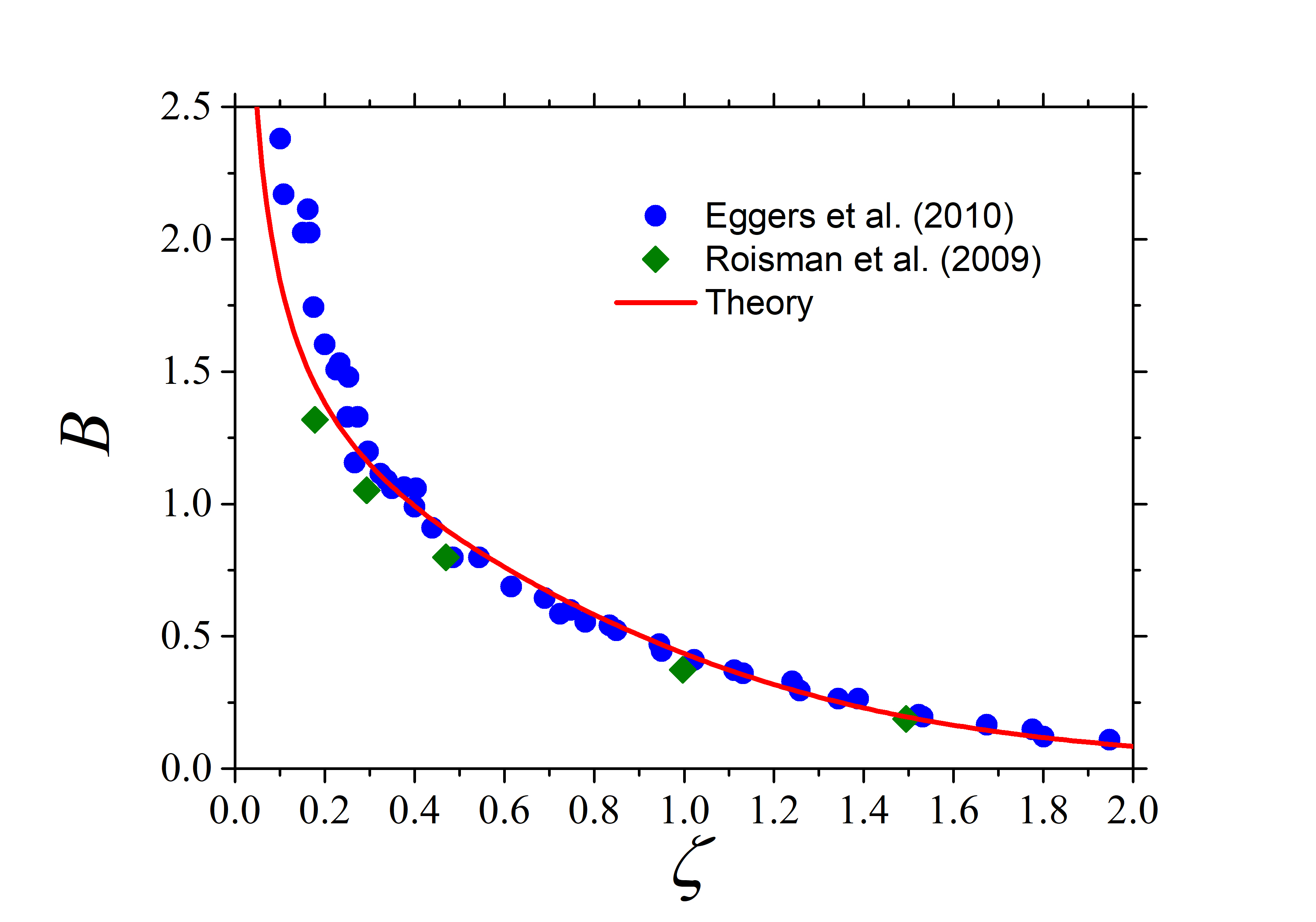}
    \caption{Theoretically predicted dimensionless pressure $B$ defined in (\ref{eq:B}) as a function of $\zeta$, in comparison with the CFD computations \cite{roisman2009inertia,eggers2010drop} of a high-speed liquid drop impact.}
    \label{fig:pinviscid}
\end{figure}

The impression radius at early times can be roughly approximated by the radius of the truncated sphere $a\approx R\sqrt{2 \zeta}$. This assumption is confirmed by the numerous experimental data \cite{rioboo2002time}. In Fig.~\ref{fig:pinviscid} the theoretical predictions for $B(\zeta)$ are compared with the CFD computations \cite{roisman2009inertia,eggers2010drop} for the initial stage of a high Reynolds number impact of a liquid drop. The agreement between the theory and the computations is rather good which means that the model accounts for the main physical players. Note that no adjustable parameters have been used in the present theory.

Integration of the ordinary differential equation (\ref{eqforK}) subject the initial conditions $K=4/3$ at $\zeta =0$, corresponding to the axial momentum of a rigid sphere, yield the following expression for $K(\zeta)$
which can be reduced to
\begin{equation}\label{eq:defK}
K=\frac{4  }{3} -  2 \int_0^\zeta \zeta B(\zeta)\mathrm{d} \zeta.
\end{equation}

\begin{figure}
    \centering
    \includegraphics[width=0.6 \textwidth]{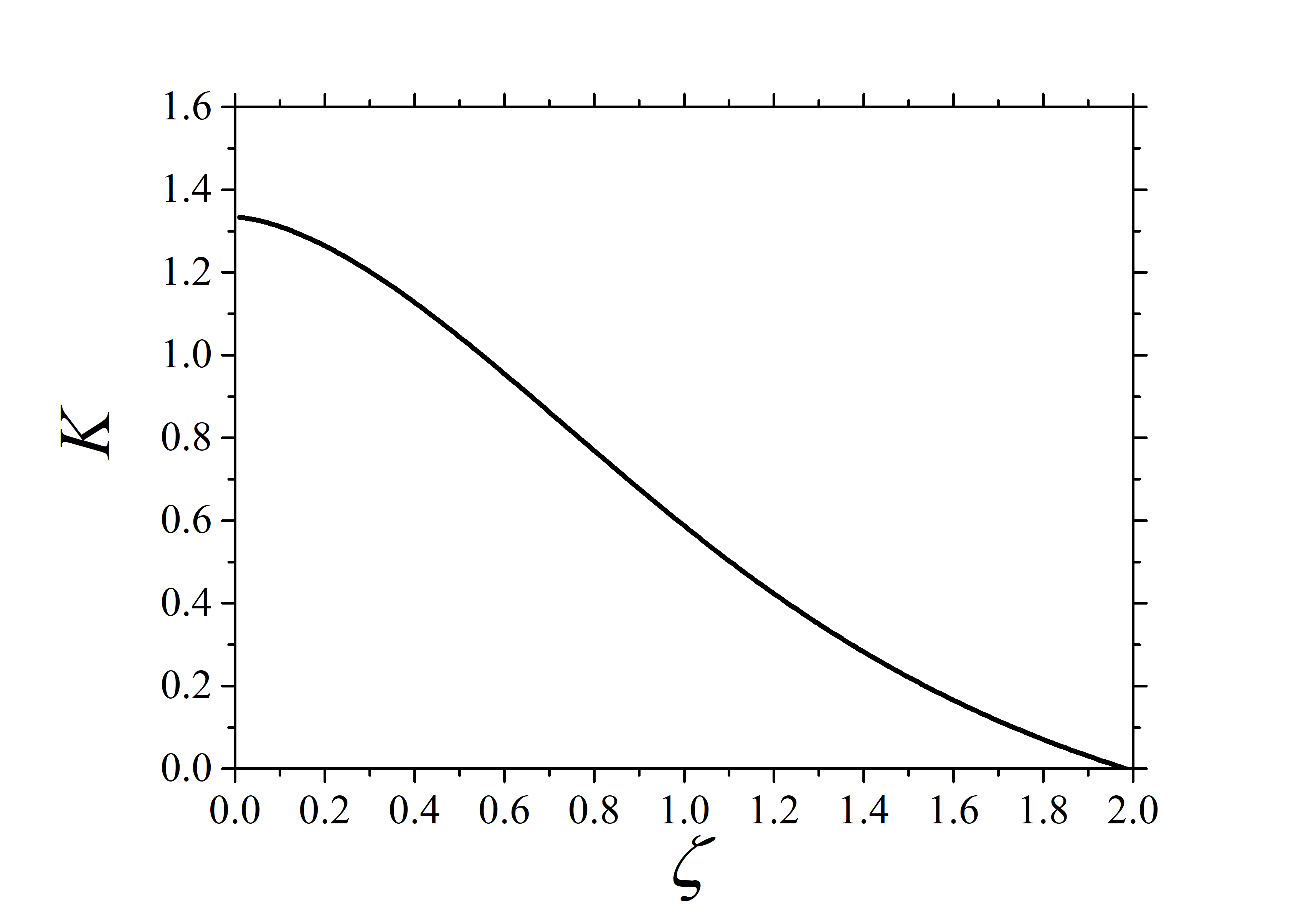}
    \caption{Theoretically predicted dimensionless axial momentum $K(\zeta)$ defined in (\ref{eq:defK}) as a function of $\zeta$.}
    \label{fig:Kinviscid}
\end{figure}

The computed values of the dimensionless axial momentum of the drop $K(\zeta)$ is shown in Fig.~\ref{fig:Kinviscid}.

Note that the present solution is valid only for relatively small values of $\zeta$. For long times $\zeta \gg 1$ the spreading of the drop is governed by a radially expanding flow in a thin lamella. Such flow has been intensively studied in the literature \cite{yarin1995impact,yarin2017collision,roisman2009inertia} and is out of the scope of this study.

\section{Equations of motion of the deforming plastic particle}

The governing equations of particle deformation can be now written in the dimensionless form using
\begin{eqnarray}
    a &=& \overline{a}(\zeta) R, \quad h = \overline h(\zeta) R, \quad t = \overline t R/U_0, \label{scales}\\
    U &=& \overline U U_0, \quad Y_0 = \rho U_0^2 \overline Y.\label{scalesb}
\end{eqnarray}

Expression (\ref{eqMomC}) in the dimensionless form is written assuming $\overline a \approx  \sqrt{2\zeta}$ with the help of (\ref{scales}), (\ref{scalesb}) and (\ref{eq:difh})
\begin{subequations}\label{eq:momY}
\begin{eqnarray}
 \overline  U\frac{\mathrm{d} \overline U}{\mathrm{d}\zeta}  &=&- \frac{2 \zeta C \overline Y}{K + 2\zeta A} .\label{eqMomCdl}\\
\frac{\mathrm{d} \overline h}{\mathrm{d} \zeta}& =& -\frac{\overline h \sqrt{2} g }{\pi\sqrt{\zeta}}-\frac{A}{\overline h},\\
       \frac{\mathrm{d}\overline t}{\mathrm{d} \zeta} &=& \frac{1}{\overline U},
\end{eqnarray}
where
\begin{eqnarray}
   A  &\equiv& \overline h \left[1  - \frac{2}{\pi}  \cot ^{-1}\left(\frac{\overline h}{\sqrt{2 \zeta}}\right)\right],\\
        B&\equiv& \frac{\overline h^2 g}{\sqrt{2}\pi \zeta^{3/2} } + \frac{1}{2} \left[\frac{A}{\overline h}+\frac{2\overline h g  }{\pi\sqrt{2\zeta}}\right]^2, \label{eq:Bdl}\\
    g &\equiv& \frac{2\zeta}{2\zeta+\overline h^2}.
\end{eqnarray}
\end{subequations}

The system of the ordinary differential equations (\ref{eq:momY}) can be integrated numerically with the help of the solution of the differential equation (\ref{eq:difh}) for $\overline h$  and of the expression for the dimensionless axial momentum of the particle (\ref{eq:defK}) subject the initial conditions
\begin{equation}
    \zeta =0, \quad \overline U=1,\quad \overline h = 2, \quad \mathrm{at}\quad \overline t = 0.
\end{equation}

The total force produced by the particle impact on the target can be now evaluated in the form
\begin{equation}\label{eq:forceZ}
    F_{z}(\zeta)\approx -\pi a^2 \sigma_{zz}=2\zeta\pi \rho R^2 U_0^2\left( B\overline U^2+\frac{C K \overline Y}{K+2 A \zeta}\right)
\end{equation}

\section{Results and discussion}

The relation of the dependence of the yield strength on the strain rate can be estimated by examining the existing values of the maximum impact force. For a perfectly plastic material the function $y(\dot \gamma)$ defined in (\ref{formY}) is constant, $y =1$. But this simplification is not sufficient for a reliable description of ice particle impact.

In this study we use the value for the static yield strength of the ice particle $Y_0=5.6$ MPa determined in \cite{tippmann2013experimentally}. The dimensionless evolution of the yield stress, defined in (\ref{formY}), is assumed in the form
\begin{equation}\label{eq:formy}
    y(\dot\gamma)= 1+\chi-\chi\exp(- \tau \dot\gamma)
\end{equation}
which corresponds to the conditions
\begin{equation}
    y(0) = 1,\quad y(\infty)\rightarrow 1+\chi,
\end{equation}
which ensures finite stresses even at very high strain rates, typical to the very early stages of particle impact $\zeta\ll 1$. The dimensionless constant $\chi$ and the characteristic time $\tau$ have to be determined from by the fitting with the experimental data.

Integration of the expression (\ref{funC}) for the function $C$ with the help of (\ref{eq:strainrateeq}) and (\ref{eq:formy}) yields
\begin{eqnarray}
    C &=& \frac{4}{3}(\chi+1) \ln \frac{2\zeta+\overline h^2}{2\zeta} +\frac{2\chi}{3} \mathrm{Ei}\left[-\frac{8\sqrt{2}\tau U_0\zeta^{3/2} \overline U}{\pi R\left(2\zeta + \overline h^2\right)^2}\right]\nonumber\\
   &-& \frac{2\chi}{3} \mathrm{Ei}\left[-\frac{2\sqrt{2}\tau U_0 \overline U}{\pi R\zeta^{1/2}}\right].
\end{eqnarray}

\begin{figure}
    \centering
    \includegraphics[width=0.6 \textwidth]{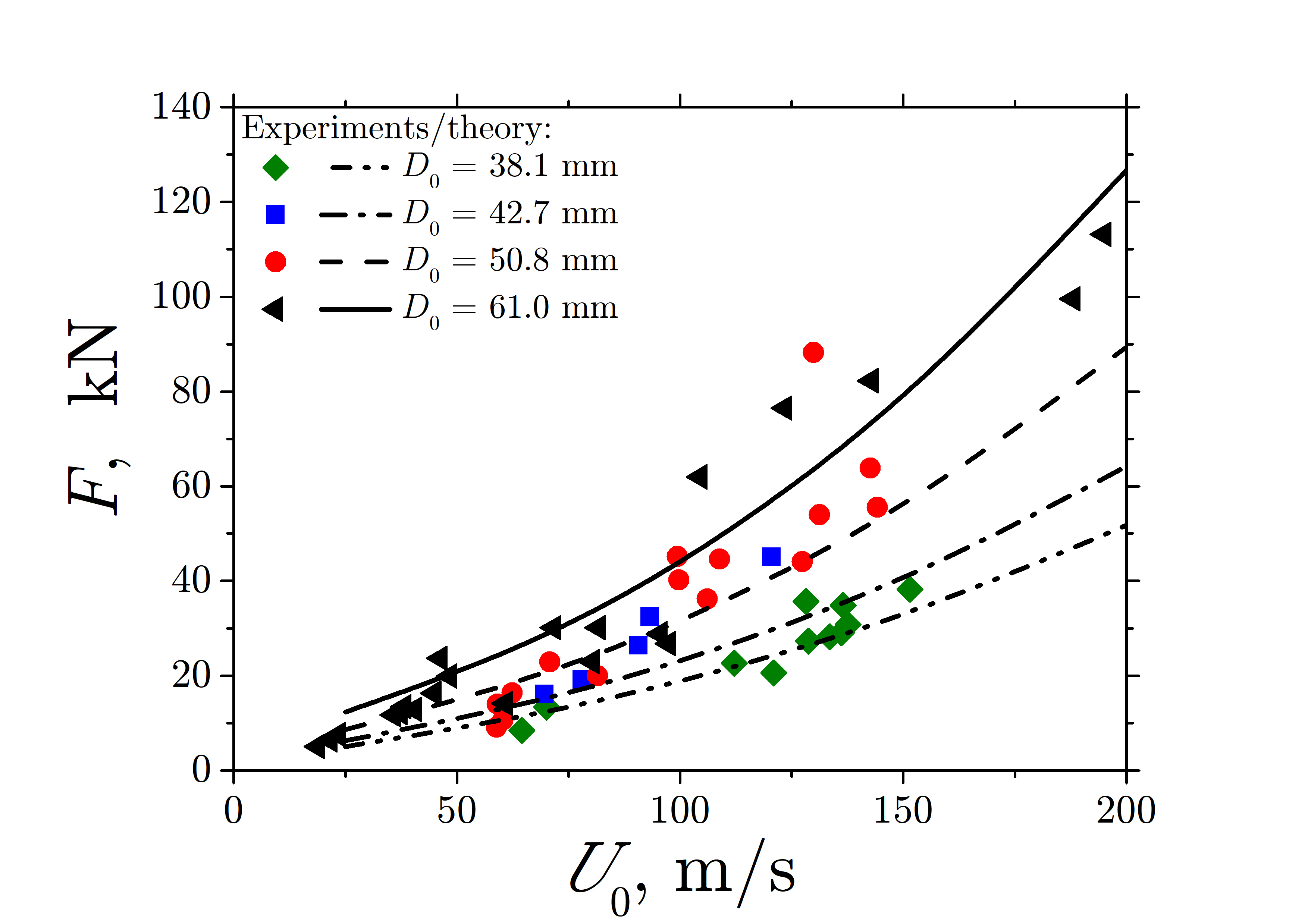}
    \caption{Comparison of the experimental data from \cite{tippmann2013experimentally} for the peak force $F$ with the theoretical predictions using fitting values $\chi=5.$ and $\tau=0.0007$ s. }
    \label{fig:maxforce}
\end{figure}

\begin{figure}
    \centering
    \includegraphics[width=0.6 \textwidth]{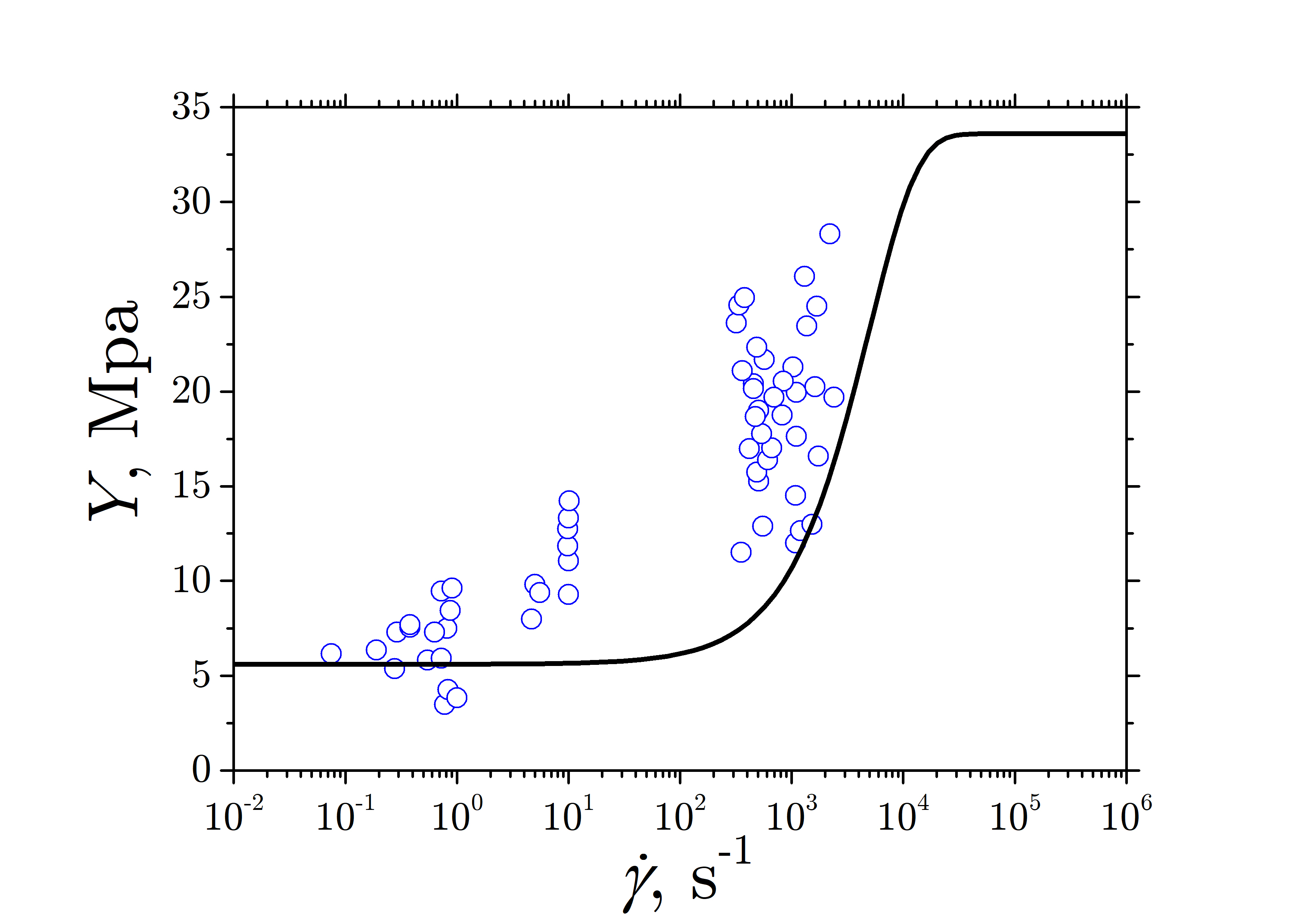}
    \caption{Comparison of the data for the yield strength $Y$ from \cite{rhymer2012force} with the assumed approximated function (\ref{formY})  used in (\ref{eq:formy}) the values $\chi=5.$ and $\tau=0.0007$ s obtained from the data for the maximum force, shown in Fig.~\ref{fig:maxforce}. }
    \label{fig:maxforce1}
\end{figure}

In Fig.~\ref{fig:maxforce} the theoretical predictions of the peak force computed using (\ref{eq:forceZ})  is plotted as a function of the impact velocity $U_0$ in comparison with the experimental data from \cite{tippmann2013experimentally}. In all the computations the fitting values $\chi=5.0$ and $\tau=0.0007$ s are used. The agreement is rather good  for various ice particle initial diameters $D_0$.  Moreover, in Fig.~\ref{fig:maxforce1} the assumed dependence of the yield strength $Y(\dot\gamma)$ on the strain rate is compared with the estimations from the full CFD computations \cite{rhymer2012force}. The values of the yield strength are of the same order despite the fact that a very simplified model is used in this study. It should be noted that the form (\ref{eq:formy}) is chosen accounting for the possibility of the derivation of the solutions for the stresses in an explicit form. We have also tried to minimize the number of the fitting parameters. Generally any form of the function $y(\dot\gamma)$ can be used in the solution for different particle materials.

\begin{figure}
    \centering
    \includegraphics[width=0.49\textwidth]{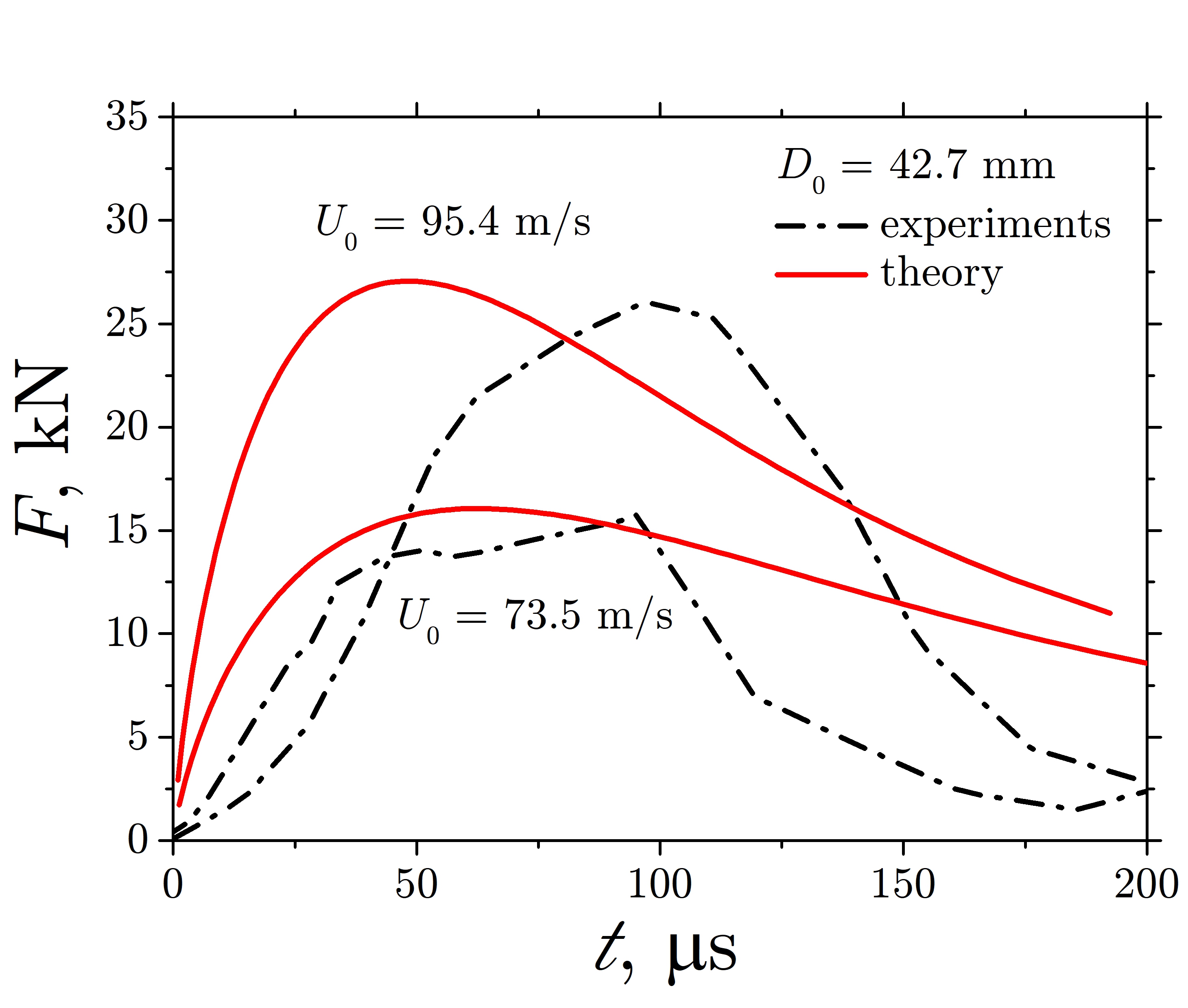}\includegraphics[width=0.49\textwidth]{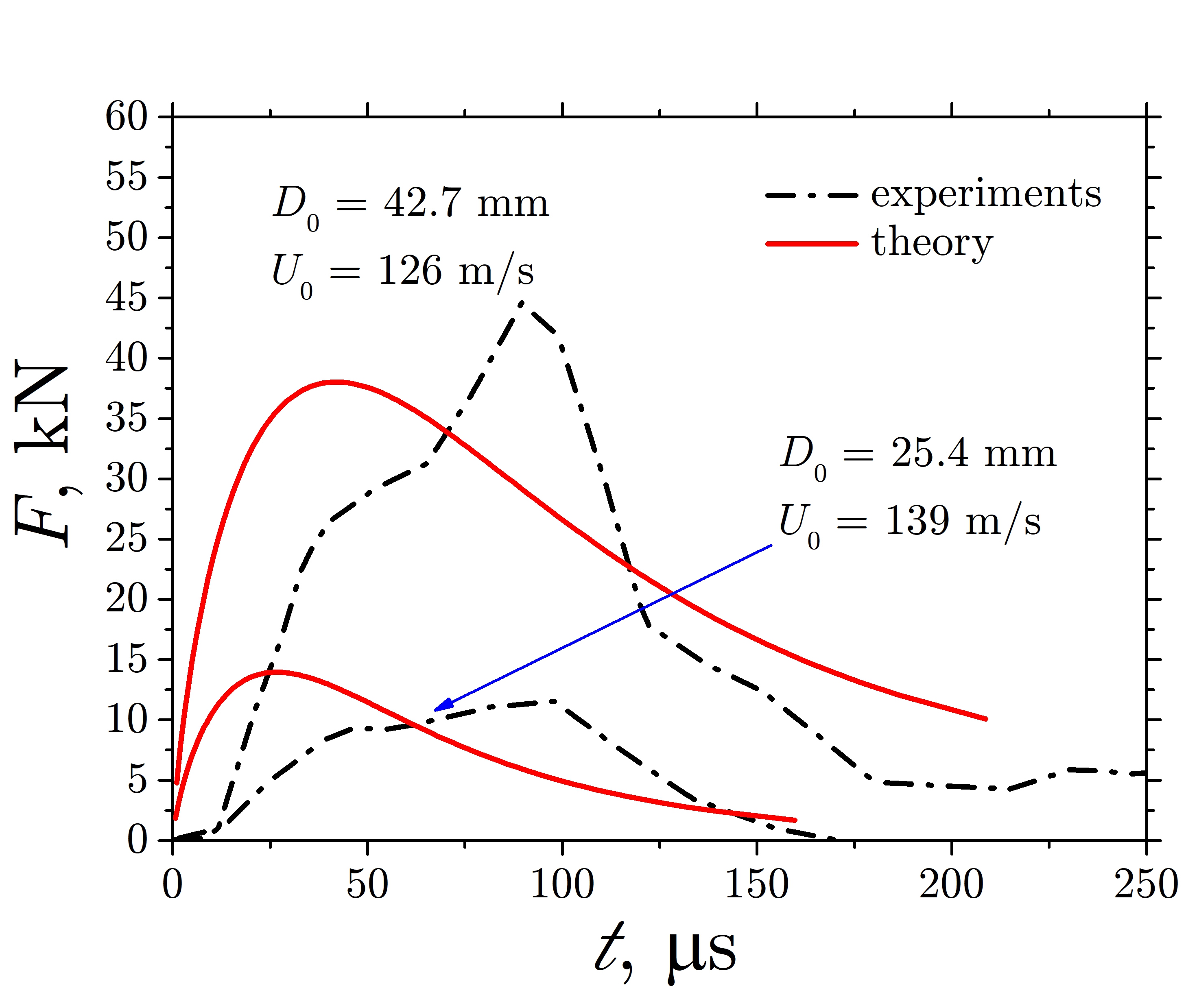}
    \caption{Theoretically predicted force evolution $F(t)$ in comparison with the experimental data from \cite{kim2000modeling}. }
    \label{fig:forcekim}
\end{figure}

\begin{figure}
    \centering
    \includegraphics[width=0.49\textwidth]{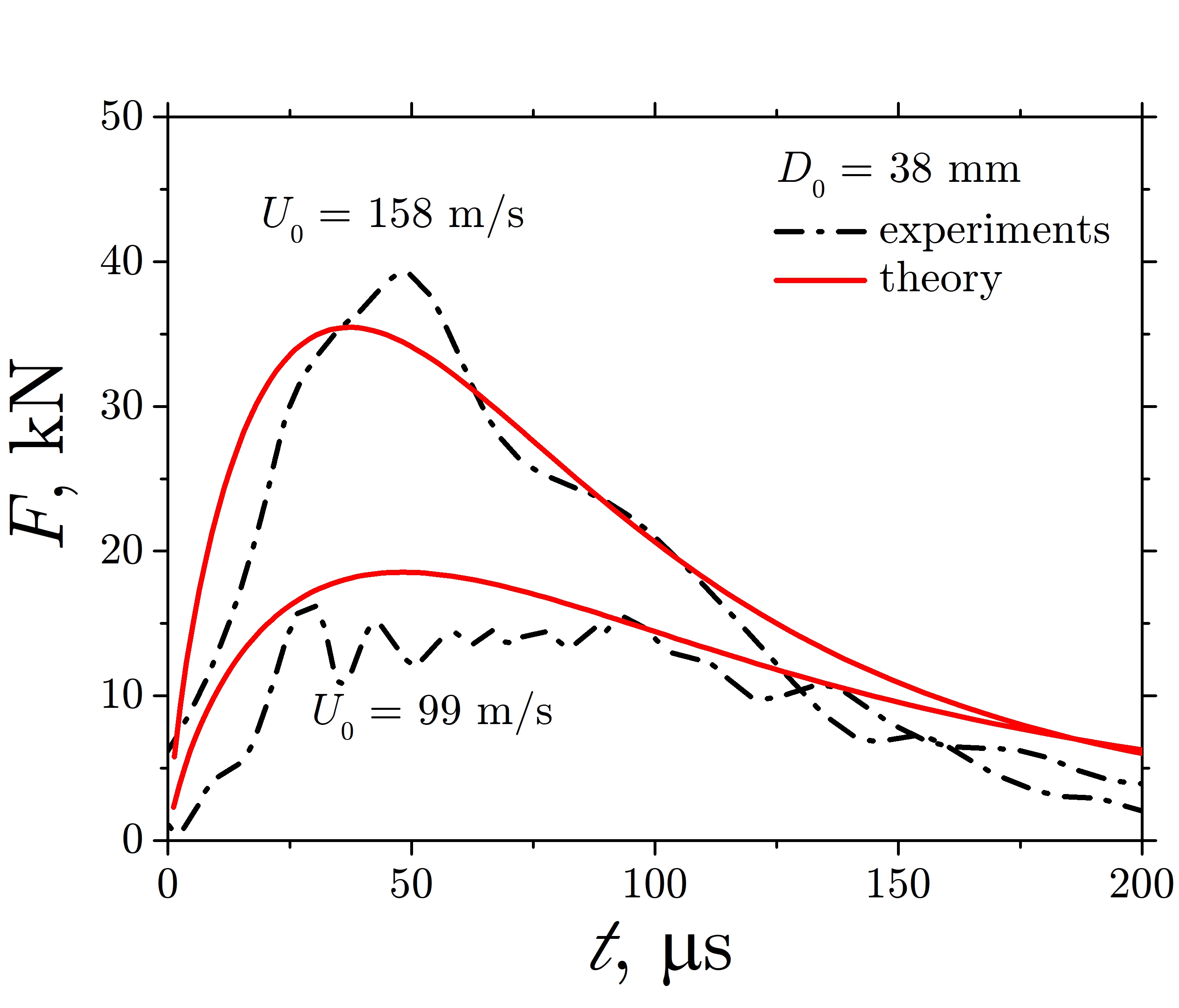}\includegraphics[width=0.49\textwidth]{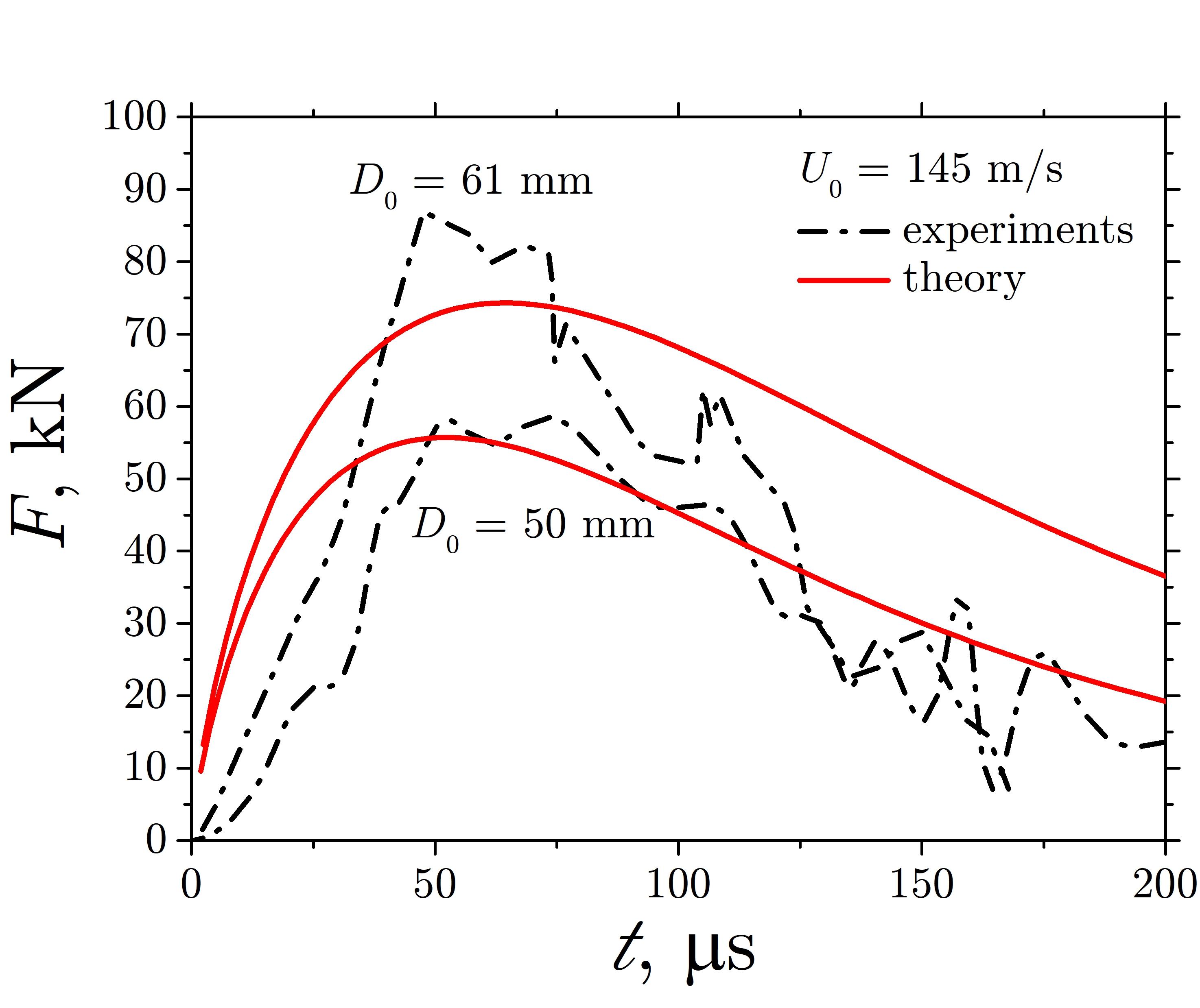}
    \caption{Theoretically predicted force evolution $F(t)$ in comparison with the experimental data from \cite{rhymer2012force}. }
    \label{fig:forcerhymer}
\end{figure}

The theoretical predictions of the force $F(t)$ computed using expression (\ref{eq:forceZ}) are shown in Figs.~\ref{fig:forcekim} and \ref{fig:forcerhymer} in comparison with the experimental data from \cite{kim2000modeling} and \cite{rhymer2012force}, respectively. In all the cases the theory allows to predict rather well the peak value and the order of magnitude  of the contact duration. In Fig.~\ref{fig:forcekim} the predicted time corresponding to the force peak is slightly underpredicted. As already mentioned in \cite{kim2000modeling} this delay can be explained by the not perfectly spherical shape of the ice particle which leads to the difficulties in the precise determination of the instant of contact. Moreover, the target used in the experiments, which is a carbon/epoxy composite panel, can be deflected by particle impact if the impact velocity is high enough. The possible target deformation and damage can be evaluated using the forces predicted by this theory, however their effect on the particle deformation is not yet accounted for.

The predictions in Fig.~\ref{fig:forcerhymer} agree with the experimental data much better than in Fig.~\ref{fig:forcekim}. The initial instant in the experiments shown in the graph in Fig.~\ref{fig:forcerhymer} is defined by the inception of the rise of the measured force.

In some cases the theory slightly overpredicts the value of the contact force at the later stages of drop deformation after the force reaches the maximum value. This overprediction can be explained by the particle fragmentation. The effect of the particle fragmentation on the value of the effective yield strength is not considered in this theoretical model.

\begin{figure}
    \centering
    \includegraphics[width=0.6 \textwidth]{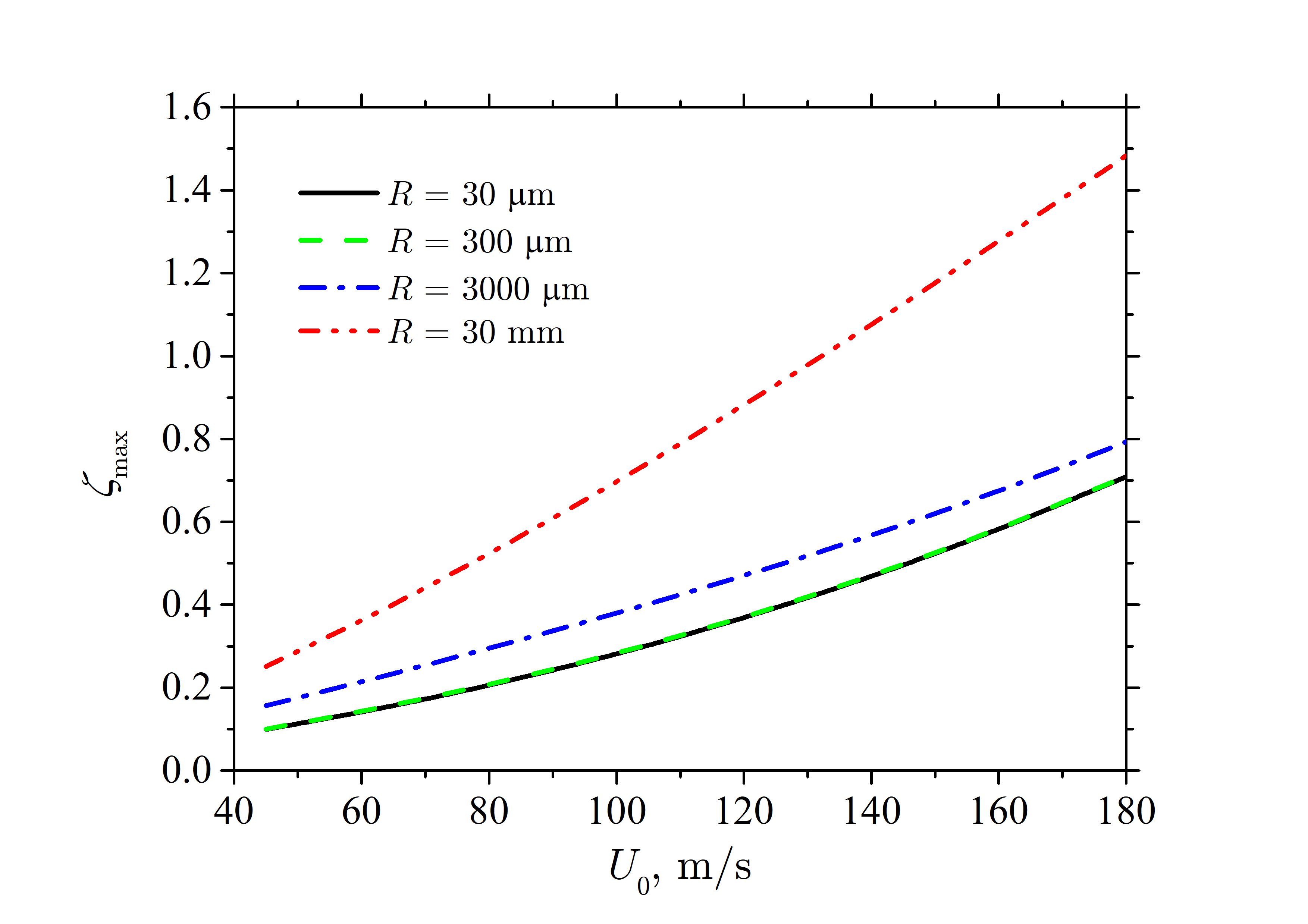}
    \caption{Theoretically predicted maximum dimensionless particle dislodging $\zeta$ as a function of the impact velocity $U_0$ for various particle radii $R$. }
    \label{fig:zMaxTheory}
\end{figure}

The value of the maximum dimensionless particle dislodging $\zeta_\mathrm{max}$ at which the particle velocity vanishes can serve as a measure of the particle total deformation. The theoretically predicted values of  $\zeta_\mathrm{max}$ are shown in Fig.~\ref{fig:zMaxTheory} as a function of the impact velocity for several particle initial radii. The value of $\zeta_\mathrm{max}$ is mainly determined by the impact velocity $U_0$. However it reduces also for smaller particles. This effect is caused by the particle hardening for higher strain rates, which are of order $\dot\gamma \sim U_0/R$. For smallest particle sizes the dependence on the radius is only minor. However, accurate experimental data with such relatively small particles are necessary for the confirmation of these predictions.

\section{Conclusions}

A theoretical model has been developed for the deformation of a spherical particle due to its normal  collision with a perfectly rigid target. The stress field in the deforming particle takes into account the hardening effects, namely the increase of the yield strength $Y$ on the local effective strain rate. This model is applied to the description of an impact of ice crystal particle.

The model is based on the potential flow field around a thin disc. It satisfies the continuity and  momentum balance equations for initial stages of impact of an spherical particle. The momentum balance equations is solved in the vicinity of the impact axis. The obtained stress at the target interface is used for the estimation of the total force.

In the limit $Y=0$ the flow is reduced to the flow in spreading inviscid drop. The theoretical predictions for the evolution of the drop height in time and for the evolution of the pressure at the substrate agree very well with the existing computational results for drop impact with very high value of the Reynolds and Weber numbers.

The dependence of the yield strength $Y$ on the local strain rate is estimated by the fitting of the theoretical predictions for the maximum force with the experiments. Finally, the model is applied for the prediction of the time evolution of the force. The theoretical predictions agree rather well with the experiments data for various particle impact velocity and initial diameter. This indicates that the theory is able not only to predict the total force but also to estimate the characteristic time of collision process and the characteristic strain rate. These results indicate that the approach in this study can be potentially applied to the prediction of the heat transfer during collision. The flow field can be used for the description of the particle fragmentation.

\section*{Acknowledgement}
This research was supported by the German Research Foundation
(Deutsche Forschungsgemeinschaft, DFG) in the framework of the SFB-TRR 150 Collaborative Research Centre, subproject A02.

\bibliographystyle{h-physrev}

\appendix
\section{Elastic region of the particle}\label{appa}
The strain tensor $\bm \varepsilon$ in the elastic region can be estimated through
\begin{equation}\label{strainvec}
   \frac{\mathrm{d}\bm \varepsilon}{\mathrm{d} t} = \bm E.
\end{equation}

The strain tensor can be expressed using (\ref{eq:uz})-(\ref{strainvec})
\begin{subequations}
\begin{eqnarray}
    \bm \varepsilon &=&\varepsilon (\bm e_r \otimes \bm e_r + \bm e_\theta \otimes \bm e_\theta - 2 \bm e_z \otimes \bm e_z),\\
    \frac{\mathrm{d} \varepsilon}{\mathrm{d} \zeta} &=& \frac{2 R a^3 }{\pi  \left(a^2+z^2\right)^2},\\  \frac{1}{R}\frac{\mathrm{d} z}{\mathrm{d} \zeta} &=& \frac{2  \cot ^{-1}\left(\frac{z}{a}\right) }{\pi }-\frac{2  a z }{\pi
   \left(a^2+z^2\right)}-1,\label{dzdzeta}
\end{eqnarray}
\end{subequations}
The impression radius $a=a(\zeta)$ is determined only by $\zeta$ and the initial shape of the particle.

The elastic-plastic boundary can be now determined from the yield condition
\begin{equation}
    G \sqrt{6\bm \varepsilon \cdot \bm \varepsilon} = Y,
\end{equation}
where $G$ is the shear modulus of ice particle. This condition at the axis yields
\begin{equation}
     \varepsilon = \frac{Y}{6 G}.
\end{equation}

\begin{figure}
\centering
\includegraphics[width=0.6\textwidth]{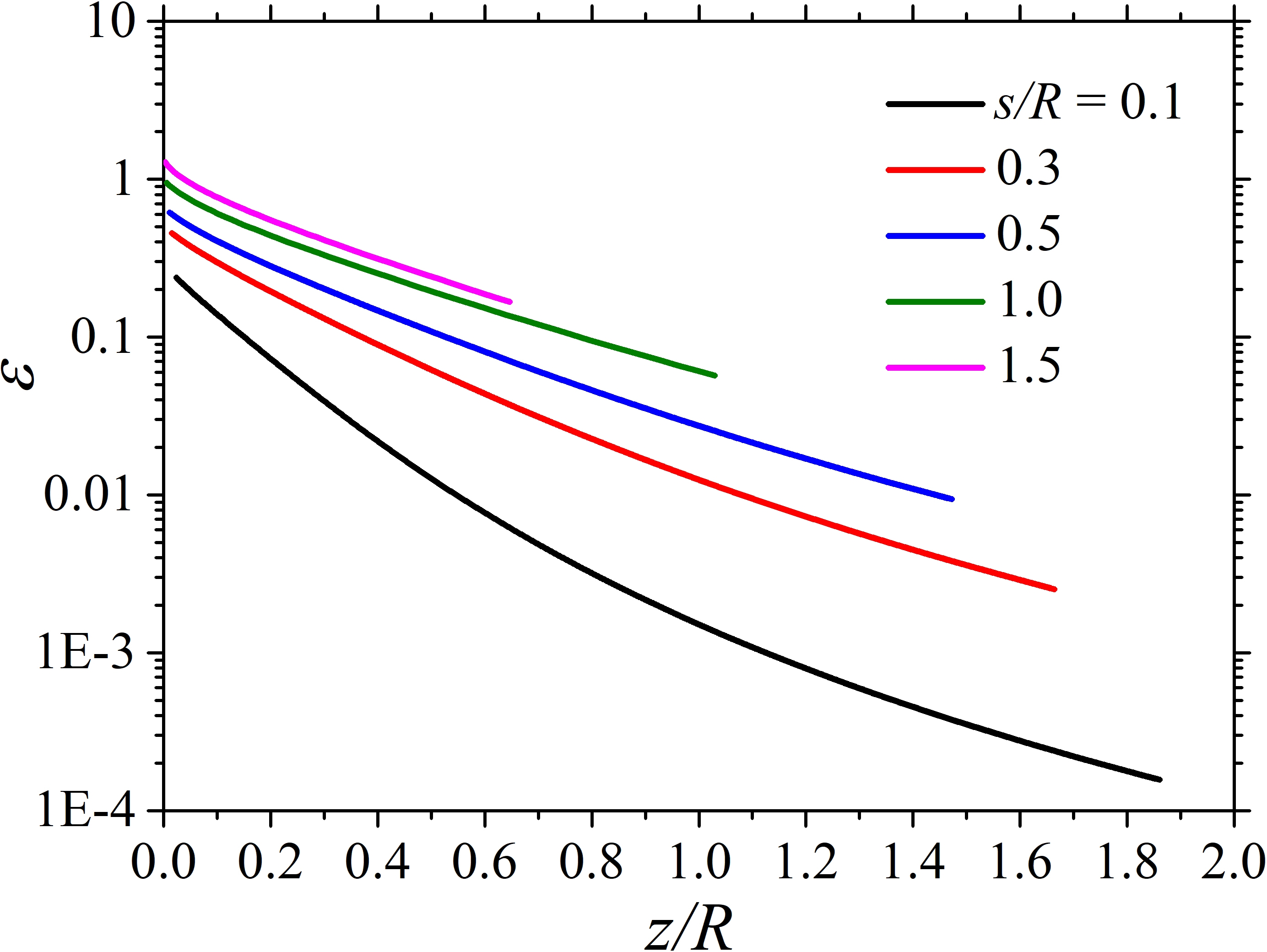}
\caption{Evolution of the strain $\varepsilon$ along the particle axis.}
\label{fig:epsilon}
\end{figure}

For a spherical particle of radius $R$, the coordinate $z=z^\star$ of the elastic-plastic boundary is obtained using the relation $a=R\sqrt{1-(1-\zeta)^2}$. In Fig.~\ref{fig:epsilon} the computed values of the strain $\varepsilon$ at the particle axis are shown for various $s$. Since for ice $Y/G\sim 10^{-3}$ already at relatively small particle displacements the particle is completely plastic.

In the further analysis for very high impact velocities leading to significant particle deformation the effect of the elasticity at the initial stage is neglected.

\end{document}